\documentstyle[12pt,epsfig,cite]{article}
\newcommand{\beq}{\begin{equation}}
\newcommand{\eeq}{\end{equation}}
\newcommand{\bea}{\begin{eqnarray}}
\newcommand{\eea}{\end{eqnarray}}

\newcommand{\gsim}{\lower.7ex\hbox{$
\;\stackrel{\textstyle>}{\sim}\;$}}
\newcommand{\lsim}{\lower.7ex\hbox{$
\;\stackrel{\textstyle<}{\sim}\;$}}

\def\ot{{\bf T}}
\def\cp{{\bf CP}}
\def\cpt{{\bf CPT}}

\begin{document}
\thispagestyle{empty}
\vspace*{-22mm}

\begin{flushright}
UND-HEP-06-BIG\hspace*{.08em}05\\
hep-ph/0604038\\

\end{flushright}
\vspace*{1.3mm}

\begin{center}
{\Large{\bf
THE SYBILS' ADVICE ON CHARM \\
\vspace{2mm}
(AND $\tau$ LEPTONS)
}}
\vspace*{19mm}

{\Large{\bf I.I.~Bigi}} \\
\vspace{7mm}

{\sl Department of Physics, University of Notre Dame du Lac}
\vspace*{-.8mm}\\
{\sl Notre Dame, IN 46556, USA}\\
{\sl email: ibigi@nd.edu}
\vspace*{10mm}

{\bf Abstract}\vspace*{-1.5mm}\\
\end{center}

\noindent 
The importance of future studies of charm and $\tau$ decays is emphasized as  
probes of New Physics: the most powerful tools are \cp~asymmetries in charm and $\tau$ decays both in partial widths and final state distributions. While these searches for \cp~violation are instrumentalized 
in the search for and hopefully future analysis of New Physics, they might also shed light on the \cp~breaking dynamics required to implement 
baryogenesis. $e^+e^-$ Super-Flavour Factories are optimally suited for the challenge, in many aspects even uniquely so. 

\vspace{2cm}

\section{Executive Summary}
\label{EXEC}

\begin{figure}[t]
\vspace{6.0cm}
\includegraphics{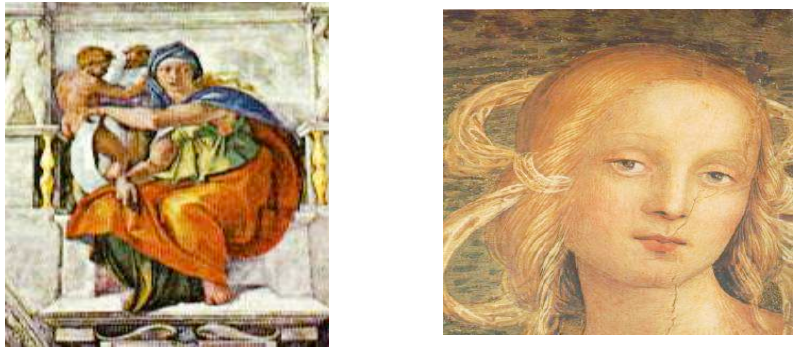}
 \caption{\it
      My IAC: La Delfica, left, and La Tiburtina, right.
    \label{IAC} }
\end{figure}

Talking about charm near Rome is particularly pleasant to me: For many of the allegories I have used over the years concerning charm have an obvious connection to 
Rome. 
\begin{itemize}
\item 
My intention: "I have come to praise C., not bury it."
\item 
My judgment:"Charm -- Come Botticelli nella Sistina"
\item 
My IAC: The Sybils -- "La Delfica" aka "Pythia" and the local one from Tivoli, "La Tiburtina", 
see Fig. \ref{IAC}. 

\end{itemize}
While the meaning of the first item is obvious, the other two need some elucidation. 
While Botticelli -- charm physics -- can match neither Michelangelo -- beauty physics --  
nor Raffaello -- kaon physics -- and therefore is often overlooked vis-a-vie the masterworks by Michelangelo in the Sistine chapel and by Raffaello nearby through an 
adverse `genius loci', he is still Botticelli, i.e. a world class artist. 
My visionary advisors emphasize four items: 
\begin{itemize}
\item 
After having found that \cp~violating phases in the quark sector can be truly large, we need to 
uncover \cp~violation in leptodynamics. The three most promising (or should I say least 
discouraging) areas are:
\begin{itemize}
\item 
neutrino oscillations; 
\item 
electron electric dipole moments (EDM's); 
\item 
$\tau$ decays. 
\end{itemize}
\item 
Baryogenesis implies the need for a `New Paradigm of \cp~Violation' -- possibly or even 
probably in leptodynamics. 
\item 
New Physics most likely induces flavour changing neutral currents: those could be (much) 
less suppressed for {\em up-} than {\em down-}type quarks. 
\item 
Charm is the only {\em up-}type quark allowing a full range of probes for New Physics: 
\begin{itemize}
\item 
Since top quarks do not hadronize \cite{RAPALLO}, there can be no $T^0- \bar T^0$ oscillations. More 
generally, hadronization, while hard to bring under theoretical control, enhances the 
observability of \cp~violation. 
\item 
As far as $u$ quarks are concerned, $\pi^0$, $\eta$ and $\eta ^{\prime}$ decays electromagnetically, not weakly. They are their own antiparticles and thus cannot oscillate. \cp~asymmetries are mostly 
ruled out by \cpt~invariance. 
\end{itemize}
\end{itemize}
In my view the justification for a Super-B factory can be and has to be based on three goals: 
\begin{enumerate}
\item 
A comprehensive and detailed analysis of $B$ (and $B_s$) transitions must be the first and foremost 
goal. 
\item 
Analyzing $\tau$ decays represent a superb second goal, since their most profound lessons might still be waiting to be learnt, and no other machine can be competitive. 
\item 
Probing charm processes constitute a still excellent third goal when one considers the envisioned 
luminosity $L \sim 10^{36}$ cm$^{-2}$ s$^{-1}$ \cite{MARCELLO} and energy flexibility coupled with the ability to study final state distributions with neutrals that is unmatched by any other set-up. 
\end{enumerate}
Thus I find it most appropriate to speak of a Super-Flavour factory. 

I would like to emphasize the following design considerations for the detector. 
\begin{itemize}
\item 
A very hermetic detector coupled with low backgrounds will be most helpful or even essential 
to control systematics in $B$, $\tau$ and charm studies, like 
$B \to \tau \tau /\tau \nu / \tau \nu X_c/\tau \nu D $, $\tau \to l \nu \bar \nu$. 
\item 
The resolution of the microvertex detector should be driven by the presumably more demanding requirements of charm physics, in particular concerning $D^0 - \bar D^0$ oscillations and 
\cp~violation there. This should benefit also searches for \cp~violation in $\tau$ decays
and in $B_s$ transitions on the $\Upsilon (5S)$ with the latter driven by $\Delta \Gamma$ 
effects. 
\item 
A {\em polarized} electron beam would be most helpful for \cp~studies in $\tau$ and 
$\Lambda_c$ decays to enhance sensitivity and control systematics.

\end{itemize}
There are several excellent reviews that the committed reader can consult to find out about details 
and further support for the statements on charm decays in this overview \cite{REVIEWS}. 

Future studies of $\tau$ and charm transitions constitutes 'hypothesis {\em generating}' rather than 
`hypothesis {\em driven}' research. Antiquity's paradigm of `hypothesis generating' analysis is 
represented by Delphi and its Pythia -- and by the nearby Tivoli with La Tiburtina. 

\section{$\tau$ Decays -- the Next Hero Candidate}

The study of $\tau$ decays, which has taught us valuable lessons on QCD, can still reveal 
the intervention of New Physics through {\em lepton flavour violating} (LFV) transitions 
\cite{HISANO,RONEY,IGONKINA} and {\em \cp~asymmetries}. 

With respect to LFV there are three classes of modes, namely (i) $\tau \to l \gamma$, 
(ii) $\tau \to l_1l_2l_3$ and (iii) $\tau \to l \nu_1 \nu_2$. While typically rates for type (i) channels  exceed those for type (ii), there are notable exceptions. Furthermore type (iii) can be probed through 
careful checks of lepton universality in $\tau$ decays \cite{PARADISI}.  Since all these types are 
forbidden in the SM, the observable rate is {\em quadratic} in the New Physics amplitude. 

\subsection{\cp~Violation}
\label{CPTAU}

As already mentioned to implement baryogenesis we need a new source of \cp~violation, and that might well be in leptodynamics driving leptogenesis as the primary effect. $\tau$ leptons provide one of the more promising areas to search for such effects: 
\begin{itemize}
\item 
The $\tau$ spin provides an important tool to enhance the experimental sensitivity. 
\item 
\cpt~constraints are less restrictive than in $\mu$ decays. 
\item 
Non-minimal Higgs dynamics, which can provide a source of \cp~violation, enjoy enhanced couplings 
in $\tau$ decays, in particular for $\tau \to \nu K\pi$, which is Cabibbo suppressed in the SM. 
\end{itemize}
Two general remarks might be of use here: (a) The sought-after \cp~asymmetry is {\em linear} in the 
New Physics amplitude, since the SM provides the other amplitude. (b) One can search for 
\cp~violation in the production of $\tau$ leptons through their EDM. Yet there one is competing against 
{\em electromagnetic} forces, in marked contrast to the situation with {\em weak} decays. Therefore 
the chances for an observable effect are better in the latter case. 

In principle one can perform a comprehensive `Fetscher-type' analysis of $\tau \to l \nu \bar \nu$ as 
for $\mu \to e \nu \bar \nu$ \cite{FETSCH}. Yet I view the channels 
$\tau ^{\pm} \to \nu K^{\pm}\pi^0/K^0 \pi^{\pm}$ as more  promising also for the observable features of its final state: such a three-body final state can exhibit \cp~violation also through asymmetries in 
the final state as discussed in general terms in Refs.\cite{MIRKES}, namely Dalitz plot asymmetries 
and/or \ot~odd distributions.  It is quite possible -- actually likely -- that asymmetries in differential 
distributions are significantly larger than in integrated rates. Furthermore they would yield more 
information on the underlying transition operator and allow consistency checks to control systematics. 

\ot~odd moments -- correlations that change sign under 
time reversal \ot~ -- allow to make efficient use of limited statistics. A \ot~odd and \cp~violating  correlation has been found in $K_L \to \pi^+\pi^- e^+e^-$ 
between the di-pion and di-lepton planes. It has been searched for in $K^+ \to \mu^+ \nu \pi^0$: 
$\langle \vec s_{\mu} \cdot (\vec p_{\mu}\times \vec p_{\pi})\rangle$. In close analogy to the latter 
one can form a \ot~odd moment in $\tau \to \nu K\pi$, if the $\tau$ spin can be exploited: 
\beq 
O_T \equiv \langle \vec s_{\tau} \cdot (\vec p_K \times \vec p_{\pi})\rangle 
\stackrel{\ot}{\Longrightarrow} - O_T
\eeq 
This would be possible most efficiently if one had the $e^-$ beam 
{\em longitudinally polarized}, since it would lead to 
the $\tau$ being produced polarized. Alternatively one can rely on a special property of 
$e^+e^- \to \gamma ^* \to \tau ^+\tau ^-$, namely that the $\tau$ pair is `spin-aligned': the polarization 
of one $\tau$ can be tagged by the decay of the other \cite{SPINALIGN}.  
Final state interactions can induce 
a non-zero \ot~odd correlation even with the dynamics \ot~conserving. Yet in 
$e^+e^- \to \tau^+\tau^-$ one can compare \cp~conjugate moments and thus isolate genuine 
\cp~violation.  

The aim should be to probe the $10^{-3}$ level; the question is whether systematic 
uncertainties can be pushed below the 1\% level without polarized beams. 

The often heard statement that practically no \cp~violation in $\tau$ decays is expected within the SM is 
not quite correct in a subtle way. As pointed out recently the `known' \cp~impurity in the 
$K_L$ wave function, which implies a corresponding impurity for $K_S$ based on \cpt~symmetry 
induces a reliably predicted asymmetry in the absence of New Physics \cite{BSTAUCP}: 
\beq 
\frac{\Gamma (\tau ^+ \to \bar \nu K_S \pi^+) -  \Gamma (\tau ^- \to \nu K_S \pi^-)}
{\Gamma (\tau ^+ \to \bar \nu K_S \pi^+) +  \Gamma (\tau ^- \to \nu K_S \pi^-)} \simeq 
2{\rm Re}\epsilon_K 
\simeq (3.27 \pm 0.12) \cdot 10^{-3} 
\eeq
The intervention of New Physics in $\tau $ decay would then modify this value. Such an 
effect does of course not exist for the similar mode $\tau ^+ \to \bar \nu K^+ \pi^0$. Comparing 
the findings in those two modes would thus provide a useful cross check on possible 
detector biases.

\section{Inconclusive $D^0 - \bar D^0$ oscillations}
\label{DOSC}

$D^0 - \bar D^0$ oscillations are a fascinating quantum mechanical phenomenon, and they 
form an important ingredient when searching for manifestations of New Physics, yet by themselves 
they represent only an ambiguous probe of New Physics. 

Oscillations can be characterized by two quantities, namely 
$x_D = \frac{\Delta M_D}{\Gamma_D}$ and $y_D =\frac{\Delta \Gamma_D}{2\Gamma_D}$.  
Oscillations  are slowed down in the SM due to GIM suppression and $SU(3)_{fl}$ symmetry. 
Comparing a {\em conservative} SM bound with the present data  
\beq 
x_D(SM), y_D(SM) < {\cal O}(0.01)  \; \; vs. \; \; 
\left. x_D\right|_{exp}  < 0.03 \; , \; \;  \left. y_D\right|_{exp} = 0.01 \pm 0.005 
\label{DOSCEXP}
\eeq 
we conclude that the search has just now begun. There exists a considerable literature -- yet 
typically with several ad-hoc assumptions concerning the nonperturbative dynamics. It is widely understood that the usual quark box diagram is utterly irrelevant due to its untypically severe 
GIM suppression $(m_s/m_c)^4$. 
A systematic 
analysis based on an OPE has been given in Ref.\cite{BUDOSC} in terms of powers of 
$1/m_c$ and $m_s$. Contributions from higher-dimensional operators with a much softer 
GIM reduction of $(m_s/\mu_{had})^2$ due to `condensate'  terms in the OPE  yield 
\beq 
\left. x_D (SM)\right|_{OPE}, \; \left. y_D (SM)\right|_{OPE} \sim {\cal O}(10^{-3}) \; . 
\label{XDYDPRED}
\eeq 
Ref.\cite{FALK} finds very similar numbers, albeit in a quite different approach. 
When evaluating the predictions in Eq.\ref{XDYDPRED} one has to distinguish carefully 
between two similar sounding questions: 
\begin{itemize}
\item 
"What are the {\em most likely} values for $x_D$ and $y_D$ within the SM?" 

My answer as given above: For both $\sim {\cal O}(10^{-3})$. 
\item 
"How large could $x_D$ and $y_D$ {\em conceivably} be within the SM?" 

My answer: One cannot rule out $10^{-2}$. 
\end{itemize}
While one predicts similar numbers for $x_D(SM)$ and $y_D(SM)$, one should keep further in mind 
that they arise in very different dynamical environments. $\Delta M_D$ is generated from 
{\em off}-shell intermediate states and thus is sensitive to New Physics, which could produce 
$x_D \sim {\cal O}(10^{-2})$. $\Delta \Gamma_D$ on the other hand is shaped by 
{\em on}-shell intermediate 
states; while it is hardly sensitive to New Physics, it involves much less averaging or `smearing' than 
$\Delta M_D$ making it thus much more vulnerable to violations of quark-hadron duality. 
{\em A similar concern applies to $\Delta \Gamma (B_s)$}. 
Observing 
$y_D \sim 10^{-3}$ together with $x_D \sim 0.01$ would provide intriguing, though not conclusive 
evidence for New Physics, while $y_D \sim 0.01 \sim x_D$ would pose a true conundrum for its 
interpretation. 

This skepticism does not mean one should not make the utmost efforts to probe 
$D^0 - \bar D^0$ oscillations down to the $x_D$, $y_D$ $\sim 10^{-3}$ level. For one we might be only one theory breakthrough away from making a 
precise prediction. Yet more importantly this challenge provides an important 
experimental validation check. A superb resolution for the $\mu$vertex detector is presumably 
essential here. 

\section{\cp~Violation with \& without Oscillations}
\label{CPV}

Most -- though not all -- factors favour dedicated searches for \cp~violation in charm transitions: 

 $\oplus$ 
Since baryogenesis implies the existence of New Physics in \cp~violating dynamics, it would be unwise not to undertake dedicated searches for \cp~asymmetries in 
charm decays, where the `background' from known physics is between absent and small: 
for within the SM the effective weak phase is highly diluted, namely $\sim {\cal O}(\lambda ^4)$, and it can arise only in {\em singly Cabibbo suppressed} transitions, where one  
expects asymmetries to reach the ${\cal O}(0.1 \%)$ level; significantly larger values would signal New Physics.  
{\em Any} asymmetry in {\em Cabibbo 
allowed or doubly suppressed} channels requires the intervention of New Physics -- except for 
$D^{\pm}\to K_S\pi ^{\pm}$ \cite{CICERONE}, where the \cp~impurity in $K_S$ induces an asymmetry of $3.3\cdot 10^{-3}$. One should keep in mind that in going from Cabibbo allowed to Cabibbo 
singly and doubly  suppressed channels, the SM rate is {\em suppressed} by factors of about 
twenty and four hundred, respectively: 
$$ 
\Gamma _{SM}( H_c \to [S=-1]) : \Gamma _{SM}( H_c \to [S= 0]) : \Gamma _{SM}( H_c \to [S= +1]) 
\simeq 
$$
\beq
1 : 1/20 : 1/400
\eeq 

$\oplus$ 
Strong phase shifts 
required for {\em direct} \cp~violation to emerge in partial widths are in general large as are the branching ratios into relevant modes;  while large final state interactions complicate the 
interpretation of an observed signal in terms of the microscopic parameters of the underlying dynamics, it enhances the observability of a signal.  

$\oplus$ 
\cp~asymmetries can be linear in New Physics amplitudes thus increasing sensitivity to the 
latter.  

 $\oplus$ 
Decays to final states of {\em more than} two pseudoscalar or one pseudoscalar and one vector meson contain 
more dynamical information than given by their  widths; their distributions as described by Dalitz plots 
or \ot~odd moments can exhibit \cp~asymmetries that can be considerably larger than those for the 
width. Final state interactions while not necessary for the emergence of such effects, can fake a signal; 
yet that can be disentangled by comparing \ot~odd moments for \cp~conjugate modes: 
\beq 
O_T(D\to f) \neq - O_T(\bar D \to \bar f) \; \; \; \Longrightarrow \; \; \; \cp~{\rm violation}
\eeq
I view this as a very promising avenue, where we still have to develop the most effective analysis tools for small 
asymmetries. 

 $\oplus$ The distinctive channel $D^{\pm*} \to D \pi^{\pm}$ provides a powerful tag 
on the flavour identity of the neutral $D$ meson. 

 $\ominus$ The `fly in the ointment' is that $D^0 - \bar D^0$ oscillations are on the slow side.

$\oplus$ Nevertheless one should take on this challenge. For 
\cp~violation involving $D^0 - \bar D^0$ oscillations is a reliable probe of New Physics: the 
asymmetry is controlled by  
sin$\Delta m_Dt$ $\cdot$ Im$(q/p)\bar \rho (D\to f)$. Within the SM both factors are small, namely 
$\sim {\cal O}(10^{-3})$, making such an asymmetry unobservably tiny -- unless there is 
New Physics; for a recent New Physics model see Ref.\cite{PEREZ}.  
One should note 
that this observable is {\em linear} in $x_D$ rather than quadratic as for \cp~insensitive quantities 
like $D^0(t) \to l^-X$.  
$D^0 - \bar D^0$ oscillations, \cp~violation and New Physics might thus be discovered simultaneously in a transition. Such effects can be searched for in final states common to $D^0$ 
and $\bar D^0$ decays like \cp~eigenstates -- $D^0 \to K_S\phi$, $K^+K^-$, $\pi^+\pi^-$ -- or 
doubly Cabibbo suppressed modes -- $D^0 \to K^+\pi^-$. In the end it might turn out that the 
corresponding three-body final states -- $D^0 \to K_S \pi^+\pi^-$, $D^0 \to K^+K^-\pi^0/\pi^+\pi^-\pi^0$ 
and $D^0 \to K^+\pi^- \pi^0$ -- allow searches with higher sensitivity. Undertaking 
{\em time-dependent} Dalitz plot studies requires a higher initial overhead, yet in the long run this 
should pay handy dividends exactly since Dalitz analyses can invoke many internal correlations 
that in turn serve to control systematic uncertainties. 
\footnote{Pythagoras' dictum "There is no royal way to mathematics" applies to fundamental physics 
as well.} 

$\oplus$ It is all too often overlooked that \cpt~invariance can provide nontrivial constraints on 
\cp~asymmetries. For it imposes equality not only on the masses and total widths of particles and antiparticles, but also on the widths for `disjoint' subsets of channels. 
`Disjoint' subsets are the decays to final states that can{\em not} rescatter into each other. Examples are 
semileptonic vs. nonleptonic modes with the latter subdivided further into those with strangeness 
$S = -1,0.+1$. Observing a \cp~asymmetry in one channel one can then infer in which other channels 
the `compensating' asymmetries have to arise. 

\subsection{A Potential New Star: $\Lambda_c$ (\& $\Xi_c$) Decays}
\label{BAR}

With the electron beam longitudinally polarized charm quarks and antiquarks would 
be produced polarized. At least some of this polarization should emerge, when those charm quarks hadronize into charm baryons, and it can be revealed through their weak decays. This would provide a 
powerful probe of \cp~violation and \ot~odd moments similar to what can happen in $\tau$ decays as 
described above. 

\subsection{Experimental Status \& Future Benchmarks}
\label{BENCH}

So far only time integrated \cp~asymmetries have been analyzed where sensitivities of order 1\% 
[several \%] have been achieved for Cabibbo allowed and once suppressed modes with two 
[three] body final states \cite{SHIPSEY}. Time {\em dependent} \cp~asymmetries (i.e. those involving 
$D^0 - \bar D^0$ oscillations) still form completely `terra incognita'. Considering the charm 
production rates achieved at the $B$ factories in particular and at FNAL I suspect the main 
limitation has been a lack of manpower rather than statistics. 

Since the primary goal is to establish the intervention of New Physics,   
one `merely' needs a sensitivity level above the reach of the SM; `merely' does not mean 
it can easily be achieved. As far as {\em direct} \cp~violation is concerned -- in partial width as well as in final state distributions -- this means asymmetries down to the 
$10^{-3}$ or even $10^{-4}$ level in  Cabibbo allowed channels and 1\% level or better 
in twice Cabibbo suppressed modes;  in Cabibbo once suppressed decays one wants 
to reach the $10^{-3}$ range although CKM dynamics can produce effects of that order 
because future advances might sharpen the SM predictions -- and one will get it along the other channels. For  
{\em time dependent} asymmetries in $D^0 \to K_S\pi^+\pi^-$, $K^+K^-$, $\pi^+\pi^-$ etc. 
and in $D^0 \to K^+\pi^-$ 
one should strive for the  
${\cal O}(10^{-4})$ and ${\cal O}(10^{-3})$ levels, respectively. 

Statisticswise these are not utopian goals considering that LHCb expects to record about 
$5 \cdot 10^7$ {\em tagged} $D^* \to D + \pi  \to K^+K^- +\pi$ events in a nominal year 
of $10^7$ s \cite{TAT}. 

When going after asymmetries below the 1\% or so one has to struggle against systematic uncertainties, in particular since detectors are made from matter. I can see three powerful weapons in this struggle: 
\begin{itemize}
\item 
Resolving the time evolution of asymmetries that are controlled by $x_D$ and $y_D$, which requires 
excellent microvertex detectors; 
\item 
Dalitz plot consistency checks; 
\item 
quantum statistics constraints on distributions, \ot~odd moments etc. 
\end{itemize}

\section{Conclusions}

Two aspects of $\tau$ decays deserve, actually require even more determined scrutiny than has been 
brought to bear up to now: lepton flavour violation (LFV) and \cp~violation. 

The observation of neutrino oscillations tells us that LFV does exist in nature. There are 
intriguing (SUSY) GUT scenarios connecting the observed $b \to s \gamma$ with 
$\tau \to \mu \gamma$ transitions and suggesting the latter to occur 
possibly at levels close to existing bounds and in any case probably within one or two orders of magnitude of them. The phenomenology of LFV in $\tau$ decays is complementary to that in 
muon decays and actually richer. While most models predict higher rates for $\tau \to l \gamma$ 
than for decays into three leptons \cite{HISANO,PARADISI} one should search also for the latter with vigour. 
The more experimental sensitivity can be achieved, the better. Finally  $\tau$ production 
in $e^+e^-$ annihilation has no practical competition from any other production process. 

We should keep in mind that we need a new \cp~paradigm to realize baryogenesis and that 
quite possibly leptogenesis might be the primary effect. This provides a more specific 
motivation to the general goal to probe \cp~violation in leptodynamics. Again $\tau$ 
decays constitute a much richer and more fertile laboratory than  muon decays. Having 
polarized electron beams leading to the production of polarized $\tau$ pairs would 
provide a very powerful though presumably not mandatory tool. It appears unlikely at 
present that one could have too much experimental sensitivity. No other setup can realistically 
compete with $e^+e^- \to \tau ^+\tau^-$ concerning \cp~studies. 

Charm is often viewed as a `has-been' quantum number: after a few absolute branching ratios have been measured accurately to provide validation opportunities to lattice QCD, there is nothing of 
substance to be learnt from them. This view overlooks that charm hadrons keep amazing us 
with features that had not been anticipated as demonstrated recently by the discovery of the 
$D_s^{**}$, $X(3872)$, $X(3940)$ and 
$Y(4260)$ resonances \cite{SHIPSEY}. Those findings have led to a re-analysis of our understanding 
-- or lack thereof -- of hadronic spectroscopy \cite{MOL,CLOSE,FOURQUARK}. 

Yet even beyond that weak charm decays might teach us important, possibly even unique lessons on New Physics. 
While it is possible to construct New Physics models that lead to \cp~asymmetries not far below present bounds and well above SM predictions, they are neither compelling nor particularly intriguing. 
Yet it behooves us to be aware of our ignorance: "We know so much about the flavour structure, 
yet understand so little." It is quite conceivable that flavour changing neutral currents are considerably stronger for up-type than for down-type quarks. Charm is the only up-type quark allowing for a full range 
of probes of New Physics through \cp~studies with and without oscillations. Charm thus might, just might provide essential support for the emerging New SM. 

Charm studies can benefit from many experimental and phenomenological advantages. 
They suffer from the drawback that charm decays in contrast to $K$ and $B$ decays are not KM 
suppressed and that oscillations proceed  at best at slow speed. Yet even on the theoretical side there 
are some advantages, namely the `dullness' of the SM electroweak phenomenology and the reasonable expectation that hadronization effects of charm can be brought mostly under theoretical control due to the comprehensive data sets accumulated by the CLEO-c and BESIII collaborations. 

We should also keep in mind that we have only recently entered a territory in charm studies, 
where one could reasonably hope to uncover New Physics -- and there are still two to three orders 
of magnitude in sensitivity waiting for the enterprising `treasure hunter'. 

One last brief comment on experimental  considerations: Doing truly superb studies of $B$ decays  has to be the paramount objective of a Super-Flavour factory. Yet it can also do superb and even unique 
$\tau$ and charm studies. One should seriously study how one can maximize the latter without in any way jeopardizing the former. One item might be to `overdesign' the $\mu$vertex detector beyond the needs of $B_{u,d}$ studies to enhance its usefulness for charm and probably also $\tau$ physics as well as for $\Delta \Gamma$ effects in $\Upsilon (5S) \to B_s^{(*)}\overline B_s^{(*)}$. 

\section{Epilogue}

Finding any signal of New Physics in high $p_{\perp}$ studies at the LHC will provide a great 
boost substantially as well as morally. Among other things it would make it mandatory to analyze 
the impact of that New Physics in heavy flavour studies. The first working hypothesis though would have to be that $B$ and $K$ decays being CKM suppressed provide the highest sensitivity to the New 
Physics -- unless one finds something rather exotic like a neutral boson decaying into, say, two jets of which only one contains charm. 

\begin{figure}[t]
\vspace{4.0cm}
\includegraphics{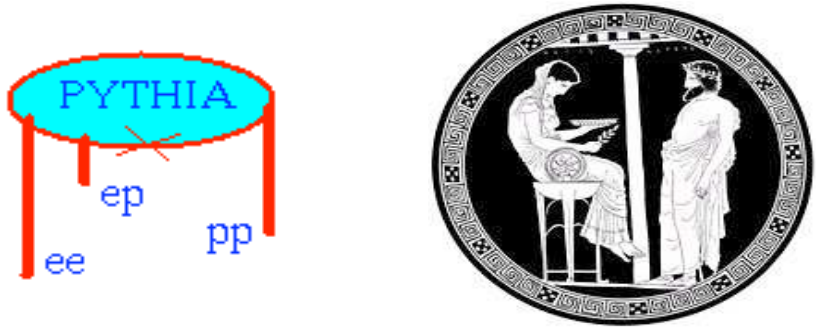}
 \caption{\it
      Two very distinct visions of Pythia.
    \label{TWOPYTH} }
\end{figure}

My message has been as specific as could be expected when coming from the Pythia on the right in 
Fig.\ref{TWOPYTH} rather the one on the left, with whom high energy physicists are more familiar, 
and when communicated by a mere mortal like me, who is not even a priest. 

\vspace{0.5cm}

{\bf Acknowledgments}

\noindent It always is a most gratifying experience to come to the Rome area for discussing  
nature's puzzles with colleagues.  This was also true this time, and I am thankful for the organizers 
of this workshop for creating this opportunity. This work was supported by the NSF under grant PHY03-55098.

\end{document}